\documentclass{emulateapj} 

\def\lsim{~\raise0.3ex\hbox{$<$}\kern-0.75em{\lower0.65ex\hbox{$\sim$}}~}
\def\gsim{~\raise0.3ex\hbox{$>$}\kern-0.75em{\lower0.65ex\hbox{$\sim$}}~}
\def\lt{~\hbox{$<$}~}
\def\gt{~\hbox{$>$}~}
\def\lbrack2{[\![}
\def\rbrack2{]\!]}

\def\mhalo{M_{\rm halo}}
\def\nhi{{N_{\rm HI}}}

\def\vcirc{{v_{\rm circ}}}
\def\v90{{v_{\rm 90}}}

\def\i0g{{I_{0\rm g}}}
\def\e0g{{E_{0\rm g}}}

\def\h2{{{\rm H}_2}}

\def\AA{{A}}

\def\km{{\rm\,km}}
\def\kms{{\rm\,km\,s^{-1}}}
\def\gcm3{{\rm\,g\,cm^{-3}}}
\def\cmsq{{\rm\,cm^{-2}}}
\def\kpc{{\rm\,kpc}}
\def\mpc{{\rm\,Mpc}}

\def\msun{{\rm\,M_\odot}}

\def\pc{{\rm\,pc}}

\def\Myrs{{\rm\,Myrs}}

\shorttitle{Velocity dispersion in DLAs}
\shortauthors{Razoumov et al.}

\begin{document}


\title{Can gravitational infall energy lead to the observed velocity
  dispersion in DLAs?}


\author{Alexei O. Razoumov\altaffilmark{1}}
\email{razoumov@ap.smu.ca}

\author{Michael L. Norman\altaffilmark{2}}
\email{mnorman@cosmos.ucsd.edu}

\author{Jason X. Prochaska\altaffilmark{3}}
\email{xavier@ucolick.org}

\author{Jesper Sommer-Larsen\altaffilmark{4,5}}
\email{jslarsen@astro.ku.dk}

\author{Arthur M. Wolfe\altaffilmark{2}}
\email{awolfe@ucsd.edu}

\author{Yi-Jung Yang\altaffilmark{6}}
\email{yjyang@asiaa.sinica.edu.tw}


\altaffiltext{1}{Institute for Computational Astrophysics, Dept. of
  Astronomy \& Physics, Saint Mary's University, Halifax, NS, B3H 3C3,
  Canada}

\altaffiltext{2}{Department of Physics, and Center for Astrophysics
  and Space Sciences, University of California, San Diego, La Jolla,
  CA 92093-0424}

\altaffiltext{3}{Department of Astronomy and Astrophysics, and
  UCO/Lick Observatory, University of California, 1156 High Street,
  Santa Cruz, CA 95064}

\altaffiltext{4}{Excellence Cluster Universe, Technische Universit\"at
      M\"unchen, Boltzmanstr. 2, D-85748 Garching, Germany}

\altaffiltext{5}{Dark Cosmology Centre, Niels Bohr Institute,
 University of Copenhagen, Juliane Maries Vej 30, DK-2100 Copenhagen,
 Denmark}

\altaffiltext{6}{Institute of Astronomy and Astrophysics, Academia
  Sinica, P.O. Box 23-141, Taipei 10617, Taiwan}


\begin{abstract}
  The median observed velocity width $\v90$ of low-ionization species
  in damped Ly$\alpha$ systems is close to $90\kms$, with $\sim 10\%$
  of all systems showing $\v90\gt210\kms$ at $z=3$. We show that a
  relative shortage of such high-velocity neutral gas absorbers in
  state-of-the-art galaxy formation models is a fundamental problem,
  present both in grid-based and particle-based numerical
  simulations. Using a series of numerical simulations of varying
  resolution and box size to cover a wide range of halo masses, we
  demonstrate that energy from gravitational infall alone is
  insufficient to produce the velocity dispersion observed in damped
  Ly$\alpha$ systems, nor does this dispersion arise from an
  implementation of star formation and feedback in our highest
  resolution ($\sim45\pc$) models, if we do not put any galactic winds
  into our models by hand. We argue that these numerical experiments
  highlight the need to separate dynamics of different components of
  the multiphase interstellar medium at $z=3$.
\end{abstract}


\section{Introduction}

Damped Ly-alpha absorbers (DLAs) provide us with a high-resolution
probe of galaxy formation processes up to redshift $z\sim5$, allowing
us to study the neutral hydrogen distribution and kinematics of young
galaxies and their surroundings on scales from several pc to several
kpc, although detailed information about the spatial extent of the
absorbing gas cannot be readily extracted from the observed velocity
line profiles. DLAs contain a large fraction of high-redshift baryons
potentially available for star formation (SF; Prochaska et al. 2005),
however, their in-situ SF efficiency appears to be a factor of 20-30
lower than in the present-day galaxies of comparable gas surface
density \citep{wolfe.06}. Even though DLAs are presumably associated
with fairly compact structures, they cover about a third of the sky,
so there is a high probability that a line of sight to a remote quasar
will contain such an absorber. Currently, just over a 1000 such
systems have been studied, and their number will undoubtedly increase
in coming years
\footnote{http://www.ucolick.org/$\sim$xavier/SDSSDLA/}.

One of the unsolved long-standing puzzles in our understanding of DLAs
is their large neutral gas velocity dispersion
\citep{prochaska.97}. The median value of $\v90$, the velocity
interval encompassing 90\% of the optical depth, in absorption lines
of low ions associated with cold neutral gas (such as SiII, NI, FeII,
etc.) is close to $90\kms$, and 10\% of all systems have
$\v90\gt210\km/s$. Observations indicate that there is virtually no
correlation between the column density $\nhi$ of the absorber and its
absorption line velocity width, with even lower column density systems
at the DLA threshold $\nhi=10^{20.3}\cmsq$ demonstrating velocity
widths up to $400\kms$. On the other hand, a correlation is observed
between the velocity widths and metallicities of the DLAs. In fact,
the correlation is remarkably tight between the equivalent width of
the \ion{Si}{2}~1526 transition (a kinematic diagnostic because the
line is saturated) and gas metallicity \citep{prochaska....07} that
shows a slope matching the local velocity/metallicity relation in
dwarf galaxies \citep{dekel.03}. These correlations may be a
consequence of a relation between the mass of the galaxies and their
metallicities \citep{wolfe.98,ledoux....06}, or an indication that
feedback from SF has a direct effect on the observed velocity
dispersion \citep{nulsen..98}.

To date all DLAs exhibit metal-line absorption
\citep{prochaska....03}, and one expects that they will all show
significant MgII equivalent widths. The opposite is not true; the
majority of MgII-selected absorbers are not DLAs. Unlike traditional
low ions, MgII traces both cold and warm neutral material, as well as
warm partially photoionized gas. Recently, \citet{murphy....07} found
a correlation between metallicity and the rest-frame MgII equivalent
width, suggesting a link between kinematics and the metal-enrichment
history of the absorber. They stressed that the absorption-line
kinematics should be viewed separately from the host galaxy
kinematics, in other words, the observed velocity widths need not
scale directly with the mass of the DLA-hosting
halos. \citet{bouche....06} used the cross-correlation between 1806
MgII absorbers and $\sim250,000$ luminous red galaxies from the Sloan
Digital Sky Survey to find that the absorber halo mass is
anti-correlated with the Mg II equivalent width, suggesting that at
least part of the observed velocity dispersion in MgII regions is
produced by supernova-driven winds and/or other feedback mechanisms
which could drive cold and warm gas efficiently out of lower mass
($\mhalo\lt10^{11.5}\msun$) galaxies. On the other hand, this
anti-correlation could also be interpreted as a transition from cold
to shock-heated gas in more massive $\sim10^{12.5}\msun$ halos
accompanied by a drop in the cumulative cross-section of neutral
clouds found in these halos \citep{tinker.07}.

At least for now, observational data are insufficient to find a
relation between the absorbing halo mass and the velocity width $\v90$
of neutral absorption in DLAs. Therefore, we should consider at least
three distinct physical mechanisms which could contribute especially
to the high-end tail of the DLA velocity distribution:

1) If cold clouds accreting onto massive galaxies are dense enough to
survive collisional heating while falling into the hot
($\sim10^6{\rm\,K}$) virialized region, they could retain a
sufficiently high neutral fraction with column densities above
$10^{20.3}\cmsq$ \citep{mcdonald.99}. Note that for this mechanism to
be viable, the accreting gas should be already sufficiently
metal-rich, as even the most metal-poor DLAs have been found to have
metallicities $Z\gsim-2.8$, i.e. it would imply a model in which SF
was ubiquitous at some time before $z=3$ in field galaxies of mass
$10^8-10^{10}\msun$. In this scenario a large fraction of metals
observed in DLAs was produced far from massive ($\gsim10^{12}\msun$)
halos where we see these metals in absorption. On the other hand, it
is also possible that the low-metallicity accreting material could mix
efficiently with the more metal-rich environment of the massive
virialized halo.

2) Conversely, DLA kinematics could be dominated by outflow velocities
of supernova-driven winds, i.e. we could see feedback directly
\citep{nulsen..98}. This hypothesis is indirectly supported by the
high detected outflow velocities in Lyman-break galaxies (LBGs), the
higher effective widths $W_{1526}$ of the SiII $1526\AA$ transition in
GRB-DLAs, and the anti-correlation between the absorbing halo mass and
the effective line width in MgII absorbers. It would also naturally
explain the metallicity-velocity correlation, since metals should be
produced in the same regions which drive the winds.

3) The apparent inefficiency of in-situ SF in DLAs, coupled with their
relatively high inferred cooling rates suggests a picture in which
very compact star-forming regions, perhaps associated with LBGs,
illuminate much more extended neutral gas structures
\citep{wolfe.06}. \citet{maller...01} used a toy model of DLA
absorption to show that a large covering factor of the cold gas in
protogalactic clumps is consistent with observations. Moreover,
numerical simulations confirm that massive halos at $z=3$ often
consist of multiple compact galaxies embedded into larger neutral
clouds. Combined outflows from stellar winds and SNe in these galaxies
could drive large chunks of surrounding neutral material to larger
galactic radii where they could pick up a higher velocity dispersion
from the local galaxy group.

Although we now have absorption data on $\sim$1000 DLAs, very few
groups have attempted to model DLA kinematics in recent years. Both
analytical and numerical models in the literature tend to rely on a
velocity dispersion put in by hand, in part because of our lack of
understanding of the physical processes of SF and feedback on
sub-galactic scales, and in part due to numerical resolution
limitations.

\citet{mcdonald.99} used an analytical model combining the
Press-Schechter formalism with a picture of individual spherically
symmetric halos consisting of multiple absorbing clouds to show that
the rate of energy dissipation corresponding to the velocity
dispersion of these clouds necessary to produce the observed line
profiles far exceeds the rate at which energy can be supplied to these
clouds by gravitational collapse and mergers of halos. They noted that
a large fraction of DLAs show multiple components in absorption
profiles of the low-ionization lines, and that this observation can be
reproduced by a model in which the missing dissipation energy comes
from supernova explosions. In fact, the energy injection rate of
$1.8\times10^{50}{\rm\,ergs\,yr^{-1}}/(10^{12}\msun)$ reproduces well
the fraction of multi-component DLAs and the overall absorption line
velocity width distribution with the median value of $\sim90\kms$.

However, their model does not provide the physical mechanism for
transferring the feedback energy to the turbulent motions in gas
clouds, apart from specifying the source of this energy. In addition,
it is a highly approximate model which assumes spherical exponential
halos consisting of discrete neutral clouds moving with a given
velocity dispersion. Moreover, the same feedback energy per unit mass
is injected into all halos, independently of their mass, an assumption
which probably puts too much velocity dispersion into small halos.

On the numerical front, \citet{nagamine...07} used cosmological
smoothed particle hydrodynamics (SPH) simulations coupled to a
phenomenological galactic wind model to compute the rate of incidence
of DLAs as a function of halo mass, galaxy apparent magnitude, and
impact parameter, for a variable strength of hydrodynamical
feedback. In their model, gas particles were driven out of dense SF
regions by hand, by assigning a momentum in random directions. The
wind mass loss rate was assumed to be twice the SF rate, and the wind
carried a fixed fraction of the SN feedback energy, corresponding to
the fixed wind velocities of $242\kms$ (weak wind) and $484\kms$
(strong wind). Depending on the strength of feedback, they found that
it evacuated gas from predominantly low-mass galaxies and increased
the cross-section and hence the rate of incidence of more massive
galaxies. In their weak feedback model $\sim10^{9.6}\msun$ halos
contributed the most to the cross-section, whereas with strong
feedback the peak shifted to $\sim10^{11}-10^{12}\msun$ halos,
depending on numerical resolution, as feedback became more and more
efficient at removing gas from low-mass systems. Although,
\citet{nagamine...07} did not analyze the velocity width statistics,
one would expect to see numerous line profiles with $\v90\gt100\kms$
in their models, due the increased incidence rate of massive galaxies.

The purpose of this paper is twofold. First, we test a hypothesis that
the observed kinematics is driven primarily by energy coming from
gravitational infall in the process of hierarchical buildup of
galaxies. This is essentially an extension of the idea put forward by
\citet{haehnelt..98} that the observed DLA line profiles are caused by
a combination of random halo motions, rotation, and infall, coupled to
the paradigm of several distinct modes of accretion onto growing
proto-galaxies \citep{dekel.06}. We use the approach developed in our
earlier paper \citep[][hereafter Paper I]{razoumov...06},
postprocessing high resolution adaptive mesh refinement (AMR)
simulations of galaxy formation with high-angular resolution radiative
transfer of UV ionizing photons. We improved our algorithm in several
ways including a better treatment of hydrodynamical heating during the
radiative transfer stage, as well as taking into account SF and
feedback. We experimented with a number of SF models, including the
standard four-criterion model of \citet{cen.92} and the two-mode SF
model of \citet{sommer-larsen..03}, both of which can be tuned to give
the observed SF rates in the absence of or with mild
feedback. Including strong feedback in our experience tends to always
suppress an ongoing SF, either unless relevant scales in the clumpy
interstellar medium (ISM) are resolved, or interaction between the
wind and the ambient medium has been turned off.

It is well known that without any special treatment the feedback
energy is quickly lost away in cooling. Several prescriptions have
been suggested to alleviate this problem; popular algorithms used in
particle-based galaxy formation models are (1) turning off cooling of
gas particles in the feedback regions
\citep{thacker.00,sommer-larsen..03,stinson.....06} and (2) using a
subresolution model for the multiphase ISM
\citep{springel.03a,nagamine...07} usually coupled with kinematic
feedback in which individual multiphase gas particles are driven out
of the star-forming regions in random or specified
directions.

Grid-based AMR simulations in which feedback energy is simply
converted into the gas thermal energy seem to be quite effective in
limiting SF, although this efficiency certainly depends on numerical
and temporal resolution. In the grid-based simulations presented in
Section~\ref{amrModels} we do not suppress cooling or use any
kinematic outflow model, in other words, we enforce a fairly
conservative approximation to the role of SF in DLA kinematics,
largely limited to conversion of some fraction of gas into stars.

In the second half of this paper (Sec.~\ref{sphModels}) we examine DLA
kinematics with TreeSPH models in which radiative cooling is
suppressed in regions surrounding active SF sites.

\section{Grid-based AMR models}
\label{amrModels}

\begin{table}
  \begin{center}
    \caption{Simulation parameters.\label{modelList}}
    \begin{tabular}{ccc}
      model & box & resolution\\
      & comoving & base$^3$+ \\
      & $\mpc/h$ & nested+AMR \\
      \hline
      N0 & 4 & $64^3+7$ \\ 
      N1 & 4 & $128^3+7$ \\
      \hline
      M0 & 8 & $64^3+7$ \\
      M1 & 8 & $128^3+7$ \\
      \hline
      L0 & 16 & $64^3+7$ \\
      L1 & 16 & $128^3+7$ \\
      \hline
      H0 & 32 & $64^3+7$ \\
      H1 & 32 & $128^3+7$ \\
      HN & 32 & $128^3+2+7$ \\
    \end{tabular}
  \end{center}
\end{table}

To accumulate the absorption line statistics, in order to compare
models to observations, we need to analyze a large cosmological
volume, at the same time attempting to resolve clumpy gas distribution
in individual galaxies. To achieve the necessary dynamical range, we
use the grid-based AMR cosmological structure formation code Enzo,
running a series of models with box sizes of $4h^{-1}$, $8h^{-1}$,
$16h^{-1}$ and $32h^{-1}$ comoving Mpc, listed in
Table~\ref{modelList}. All of our models have the base grid size of
$128^3$, with up to seven additional levels of refinement everywhere
in the volume. Each level of refinement features twice the spatial and
8 times the mass resolution of the previous level. In order to check
resolution effects, for each box size we also ran one $64^3$ base grid
model with up to seven levels of AMR. In addition, we ran a
high-resolution simulation HN of the most massive halo in the
$32h^{-1}\mpc$ model H1. For this model we traced the positions of all
dark matter (DM) particles found at $z=3$ inside the spheres of radii
$2h^{-1}\mpc$ and $4h^{-1}\mpc$ centered on the target halo back to
the initial redshift $z=99$, and regenerated the initial conditions
(ICs) with 2X the grid and 8X the mass resolution inside the region
which contributed all DM particles to the $4h^{-1}\mpc$ sphere, and
with 4X the grid resolution and 64X the mass resolution inside the
region which contributed the DM particles to the $2h^{-1}\mpc$
sphere. Since the base grid resolution in this run is $128^3$, the
effective resolution for the halo is $512^3$, with 7 additional levels
of AMR, resulting in the same physical ($200\pc$) and mass resolution
as the entire $8h^{-1}\mpc$ volume in model M1.

Star formation was modeled with discrete stellar particles using the
prescription of \citet{cen.92}, with the overdensity threshold
$\delta_{\rm cr}=100$, the star formation efficiency $\epsilon_{\rm
  sf}=0.1$, and the stellar particle mass $M_*=10^6\msun$. Whenever
the conditions for SF were met and there was enough gas in a cell to
form at least one stellar particle, this gas was converted into
stars. Although a stellar particle was created instantaneously, the
actual star formation and feedback was modeled on the local dynamical
timescale. At each timestep, a $\epsilon_{\rm th}=10^{-5}$ fraction of
the rest-mass energy of the stars produced during that time interval
was injected locally as thermal energy.


\subsection{UVB radiative transfer}
\label{rt}

Our galaxy formation models include a non-equilibrium ionization
network for hydrogen and helium, in the presence of a uniform
\citet{madau..99} UVB. The output of these simulations at $z=3$ is
then postprocessed iteratively with the UVB radiative transfer code
FTTE \citep{razoumov.05} developed for nested grids, until an
equilibrium position of hydrogen and helium ionization fronts is
found, with the angular resolution of 192 bins over the $4\pi$ sphere,
using the same UVB as in the hydrodynamical run. This algorithm is
similar to the one we used in Paper I, but with an improved treatment
of hydrodynamical heating. In Paper I we added a location-specific
hydrodynamical heating term which would keep the temperature of each
cell constant during the iterative radiative transfer stage by exactly
balancing radiative cooling, provided that the radiation field in that
cell stays the same as in the hydrodynamical calculation. Recently we
found that due to numerical instabilities this approach had a tendency
to overestimate the neutral fraction in the shock-heated regions of
very massive halos where the radiative cooling timescale is very
long. Now we simply do not update temperature in shock heated regions
with $T\gt T_{\rm cr}$, where $T_{\rm
  cr}=3\times10^4{\rm\,K}$. Although heating by active galactic nuclei
via helium photoionization could take the temperature above
$3\times10^4{\rm\,K}$, in practice we found that our results are not
sensitive to the exact choice of $T_{\rm cr}$, as long as it stays in
the $2\times10^4-5\times10^4{\rm\,K}$ range, since shock-heated
regions have fairly narrow boundaries.

\subsection{Spectrum generation}

To generate and process artificial spectra, we follow the procedure
outlined in Paper I, with one important modification. The principal
line width diagnostic used throughout this paper is the velocity
interval $\v90$ encompassing 90\% of the optical depth in the
line. This diagnostic is dominated by clouds with large optical depths
-- its more detailed discussion and the comparison to the equivalent
width can be found, e.g., in \citet{prochaska....07}. To measure
$\v90$, we need an unsaturated line which also has a sufficient
oscillator strength to stand out above the noise. In observations such
a line is picked by hand from a set of low ionization lines associated
with the absorber. In numerical models we deal with a very large
number of sightlines to accumulate the statistics and therefore need a
robust algorithm to pick up the suitable line automatically. One
solution is to adjust the line strength by hand, so that the optical
depth in the line center is always $\tau_{\rm c}\sim2-3$. Here we
simply compute $\v90$ prior to applying the noise to the
spectrum. This approach nevertheless gives us the right measurement
since $\v90$ is not sensitive to the choice of the specific line
provided that all of its components can be seen above the noise.

\subsection{Results from AMR models}

\begin{figure*}
  \epsscale{1.1}
  \plotone{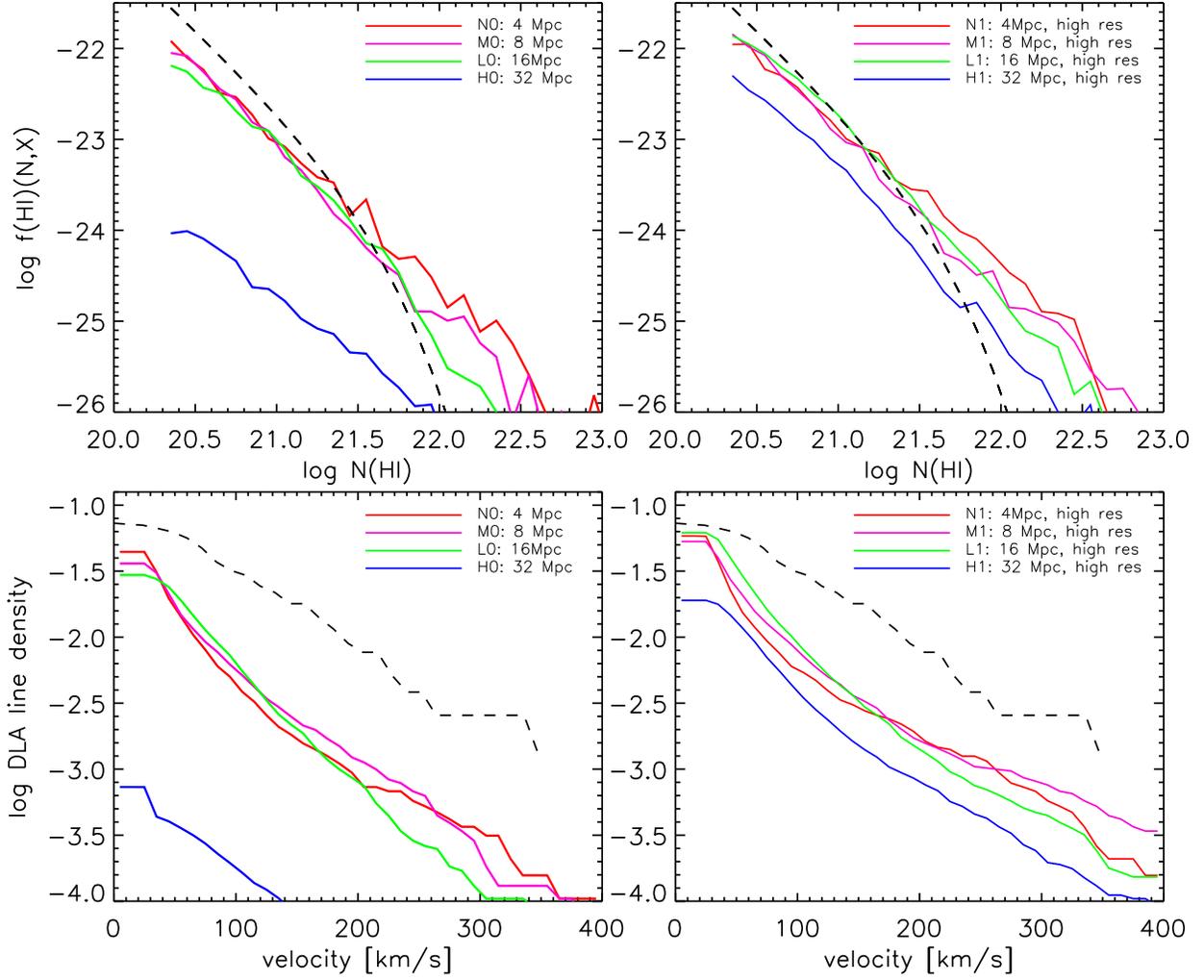}
  \caption{HI column density frequency distribution $f(N,X)$ (top
    panels, see text for details) and line density $\ell_{\rm DLA}(X)$
    of DLAs with the Si II velocity width higher than $v_{\rm Si II}$
    vs. $v_{\rm Si II}$ (lower panels) for $4h^{-1}\mpc$ (red lines),
    $8h^{-1}\mpc$ (magenta lines), $16h^{-1}\mpc$ (green lines) and
    $32h^{-1}\mpc$ (blue lines) models, at $z=3$. The panels on the
    left are for the $64^3+7$ (i.e., 7 additional levels of
    refinement) models, the panels on the right are for the higher
    resolution $128^3+7$ models. The dashed lines are fits to the
    observed distributions at $z=3$ \citep{prochaska..05}.}
  \label{colVel-boxSize}
\end{figure*}

Figure~\ref{colVel-boxSize} provides a quick glance at our
results. The top panels show the HI column density frequency
distribution $f(N,X)$ defined such that $f(N,X)dNdX$ is the number of
DLAs in the intervals $[N,N+dN]$ and $[X,X+dX]$, where $N$ is the HI
column density, and $dX$ is the ``absorption distance'' interval

\begin{equation}
dX={H_0\over H(z)}(1+z)^2dz.
\end{equation}

\noindent
The bottom panels of Figure~\ref{colVel-boxSize} show the line density
$\ell_{\rm DLA}(X)$, i.e. the number of DLAs per unit absorption
distance with the velocity width higher than $\v90$, for $64^3+7$ (on
the left) and $128^3+7$ (on the right) models,
respectively. Figure~\ref{colVel-boxSize} gives the same distributions
as Figures 4 and 5 in Paper I, with several differences. First, our
latest models use consistently higher grid resolution, across the
wider range of volumes from $4h^{-1}$ to $32h^{-1}\mpc$. Next, we
calculate the neutral hydrogen fraction in shock-heated regions more
accurately. Finally, we account for conversion of gas into stars, as
well as feedback, albeit in a conservative form. Only model M1 has the
same grid and mass resolution as one of the runs (C2) in Paper I,
however, the above-mentioned changes yield slightly lower HI column
densities across the entire $\nhi$ range.



At low column densities ($10^{20.3}-10^{20.6}\cmsq$) we find fewer
absorbers than observed, with the maximum discrepancy factor of
$\sim1.6$ at the DLA threshold. Part of this discrepancy may be
explained by a Malmquist bias in the observational result
\citep{omeara.....07}. On the other hand, our models produce
consistently more high-column density absorbers, missing the turn-off
at $\nhi\gsim10^{21.5}\cmsq$. This behaviour is not surprising as (1)
we do not include the formation of $\h2$ molecules on dust grains,
which is believed to be the major sink for neutral hydrogen atoms at
higher densities, and (2) we do not account for gas clumping on scales
below $100\pc$. Apart from the uncertainties of $\h2$ formation at
non-zero metallicities, putting this physics in by hand into our
current models is not yet practical, as we do not resolve the
$\sim1-10\pc$ scale of individual cold clouds in which these molecules
form. Note that $\nhi=10^{21}\cmsq\approx8\msun{\rm\,pc^{-2}}$ is also
a critical surface density threshold for the formation of the cold
phase \citep[e.g.,][]{schaye04}. Our limited numerical resolution in
effect prevents conversion of gas into the cold phase which would
otherwise trigger gravitational instability, i.e. further collapse of
densest cores leading to much smaller cross-sections. In addition,
once the multiphase medium develops, hydrodynamical feedback from OB
winds and supernovae in such a clumpy medium is likely to lead to
compression of HI regions into thinner shells again producing lower
cross-sections. Gas clumping on these small scales is expected to
lower the rate of incidence of clouds forming $\h2$ which probably
explains why the observed transition between the HI and $\h2$ column
density distributions at $z=0$ \citep{zwaan.06} corresponds to fewer
absorbers ($\log f(N,X)\sim-26$ at $\nhi=10^{22}\cmsq$ ) than what is
predicted in our current models ($\log f(N,X)\sim-24.5$ at
$\nhi=10^{22}\cmsq$).

\begin{figure}
  \epsscale{1.1}
  \plotone{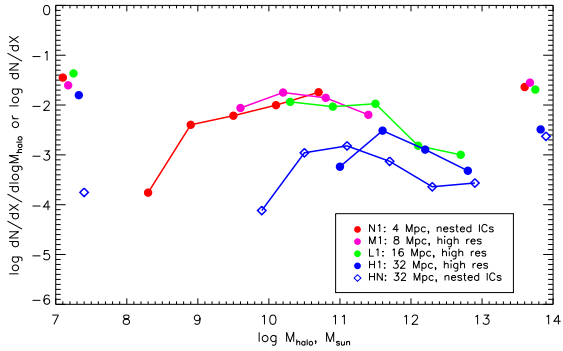}
  \caption{Connected points show the differential line density $dN/(dX
    d\log\mhalo)$ of DLAs as a function of halo mass. Isolated points
    on the left ($\mhalo<10^{7.4}\msun$) and on the right
    ($\mhalo>10^{13.6}\msun$) show the cumulative line density $dN/dX$
    of intergalactic and halo DLAs, respectively.}
  \label{incidenceMass}
\end{figure}

\begin{figure*}
  \epsscale{1.1}
  \plotone{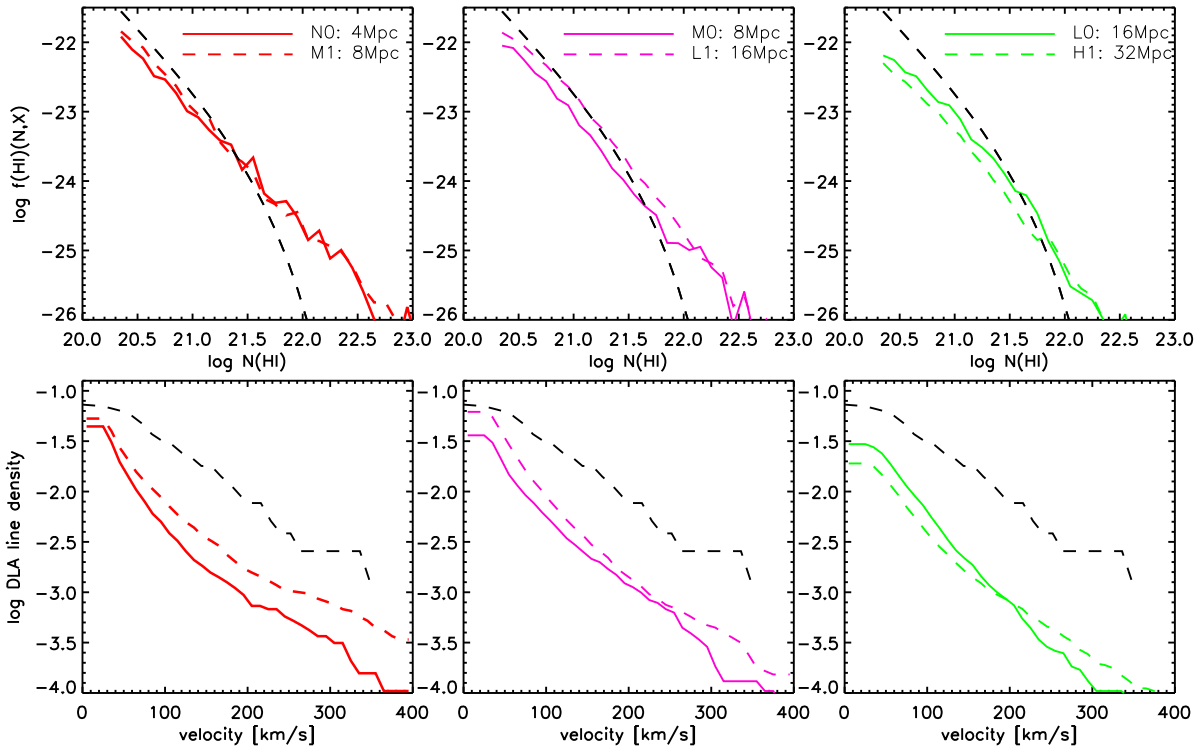}
  \caption{Same as Fig.~\ref{colVel-boxSize} except that now each
    panel shows two simulations of same grid and mass resolution but
    different volumes.}
  \label{colVel-sameRes}
\end{figure*}

\begin{figure}
  \epsscale{1.1}
  \plotone{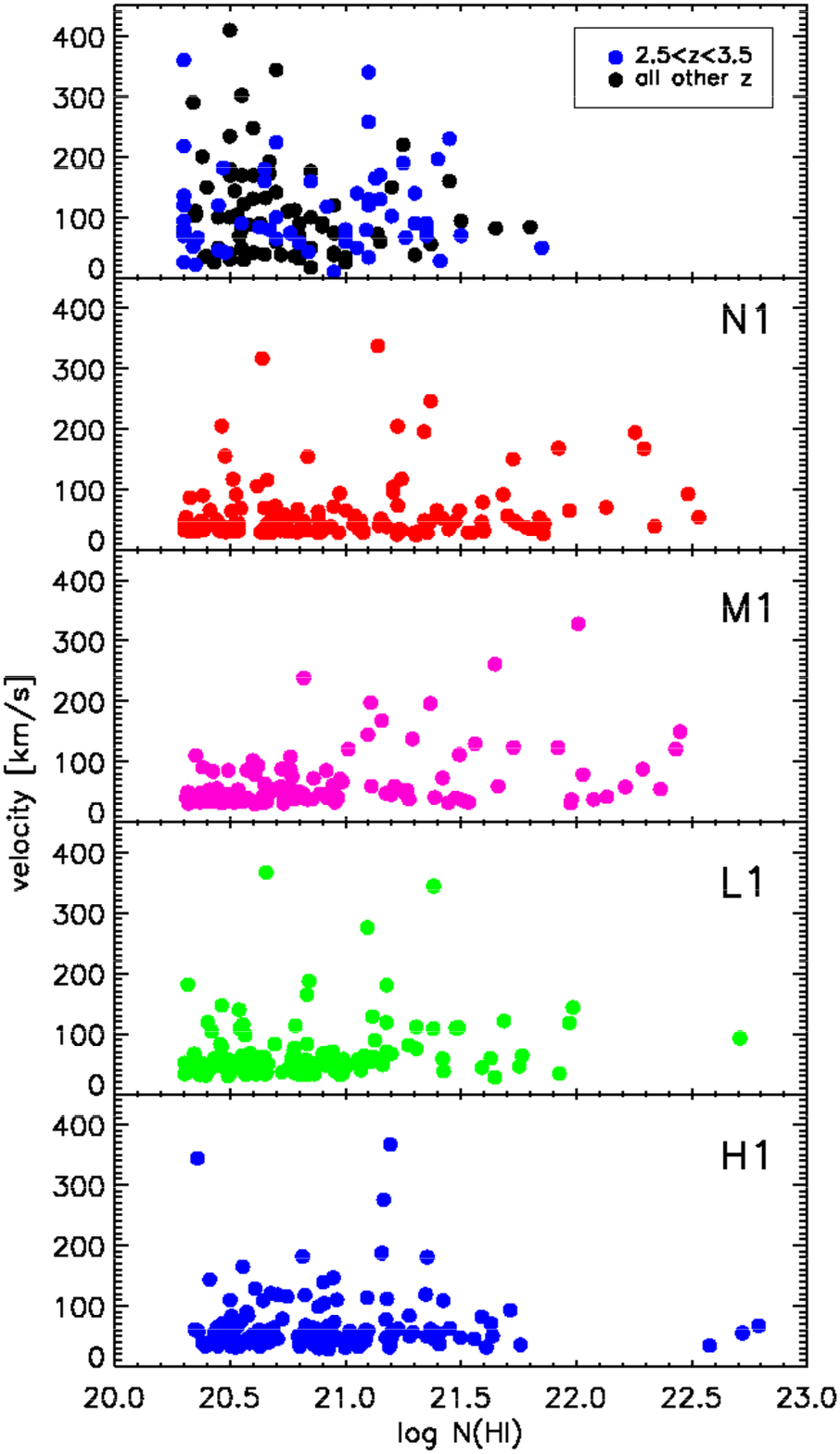}
  \caption{HI column density vs. velocity scatter plot for a sample of
    135 observed DLAs from the SDSS DLA survey \citep[][top
    panel]{prochaska..05} and for 135 randomly picked DLAs from models
    N1, M1, L1 and H1.}
  \label{scatter-colVel-boxSize}
\end{figure}

We see the same trend found in Paper I that higher grid resolution
partially alleviates the kinematics problem producing a larger
fraction of $\v90\gsim70\kms$ absorbers. However, the median value of
$\v90$ in the resolved models N1, M1 and L1 ranges from 39 to
$53\kms$, well below the observed value of $90\kms$. In model H1 the
median $\v90$ is $64\kms$, but this model is clearly not resolved as
it misses a large fraction of absorbers with $\mhalo\lsim10^{11}\msun$
(Fig.~\ref{incidenceMass}). Similarly, going to larger computational
volumes at a fixed grid and mass resolution
(Fig.~\ref{colVel-sameRes}) accounts for progressively more massive
halos and in smaller boxes tends to produce a larger fraction of
high-velocity systems. The low median value of $\v90$ is further
illustrated in the HI column density vs. velocity scatter plot
(Fig.~\ref{scatter-colVel-boxSize}) in which we compare a quasi-random
sample of 135 observed DLAs to a sample of 135 systems picked randomly
from each model.

\begin{figure}
  \epsscale{1.1}
  \plotone{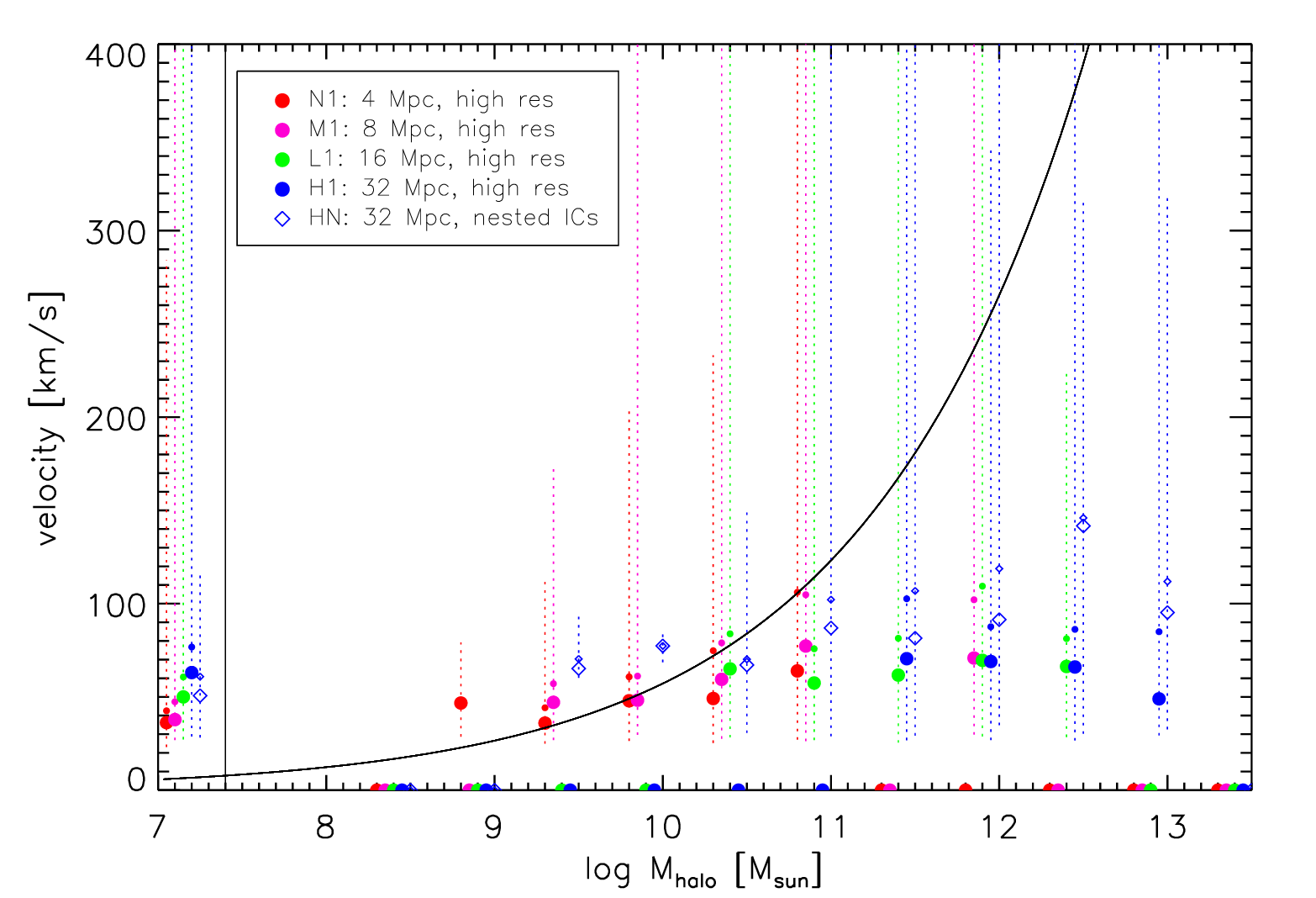}
  \caption{Median (large symbols) and mean (small symbols) $\v90$
    velocity widths for DLA sightlines which pass through virial radii
    of DM halos with $\mhalo>10^8\msun$, vs. $\mhalo$, in models S1
    (red filled circles), M1 (magenta filled circles), L1 (green
    filled circles), H1 (blue filled circles) and HN (blue open
    diamonds). Dotted lines show the full range of absorption
    velocities in each mass bin, separately for each model. The thick
    solid line is the z=3 circular velocity. Data points on the left
    ($\mhalo<10^{7.3}\msun$) show the velocity widths for
    intergalactic DLAs, i.e. absorbers not associated with any halo.}
  \label{velMass}
\end{figure}

Let us now look at how halos of different masses contribute to the
overall kinematics. Figure~\ref{velMass} shows the $\v90$ velocity
widths for those DLA sightlines which pass through the virial radii of
all DM halos in each model. These DLAs do not encompass our entire
simulated sample, as we see a substantial fraction of
$\nhi\gt2\times10^{20}\cmsq$ systems in filaments or intergalactic
neutral clouds not associated with any particular DM halo
(Table~\ref{intergalacticDLAs}). This fraction varies with resolution
and box size but is fairly large in all models, except for the nested
ICs simulation of the massive halo HN in which there are very few
intergalactic DLAs. Most of the volume in this model is not resolved
to have DLA-type absorption, except for the $2\mpc$ refined
region. This region corresponds to a higher density peak in the
initial density distribution in which structures started to form
earlier, with less gas left over in the diffuse filamentary form,
which explains its very high fraction of DLAs associated with halos
(93\%).

Note that at much higher numerical resolution a lot of these
intergalactic absorbers would accrete onto low-mass halos or form more
compact filamentary structures with a smaller overall cross-section,
so that the values given in Table~\ref{intergalacticDLAs} should
probably be regarded as upper limits to the average fraction of
intergalactic DLAs at $z=3$.


\begin{table}
  \begin{center}
    \caption{Fraction of intergalactic DLAs.\label{intergalacticDLAs}}
    \begin{tabular}{clll}
      box, $\mpc/h$ & low res & high res & nested\\
      \hline
      4 & 44.2\% & 57.6\% &\\
      8 & 49.7\% & 39.6\% &\\
      16 & 43.9\% & 54.8\% &\\
      32 & 26.0\% & 62.8\% & 7.0\%\\
    \end{tabular}
  \end{center}
\end{table}

The scatter in Figure~\ref{velMass} is quite large. Apart from the
resolution effects, there is a correlation between the observed median
$\v90$ and the circular velocity, although this dependence has a
flatter slope than the $v_{\rm obs}\sim0.6\vcirc$ found by
\citet{haehnelt..98}. Moreover, we see that this relation might break
at higher masses where individual halos consist of multiple
components, and unless these components are resolved the median of the
observed velocity widths can be as small as $\sim0.2\vcirc$.

\begin{figure}
  \epsscale{1.1}
  \plotone{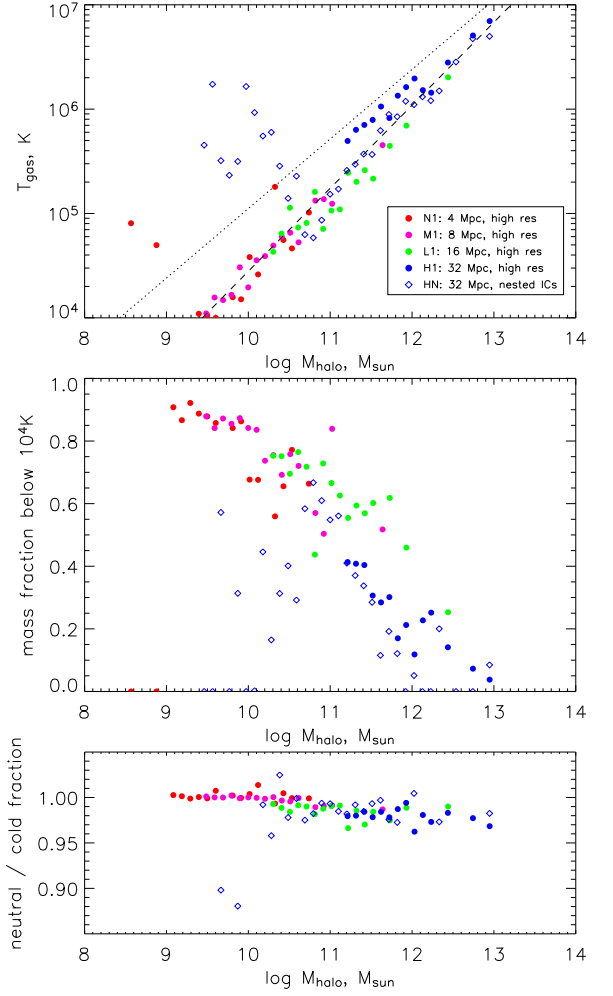}
  \caption{Mass-weighted average gas temperature (top panel), fraction
    of cold ($T\lt10^4{\rm\,K}$) gas (center panel), and ratio of the
    mass-weighted average hydrogen neutral fraction to the fraction of
    cold gas (lower panel) inside virial radii of all halos in a given
    mass bin, vs. that mass, for model S1 (red filled circles), M1
    (magenta filled circles), L1 (green filled circles), H1 (blue
    filled circles) and HN (blue open diamonds). The dotted line in
    the top panel shows the virial temperature at $z=3$, while the
    dashed line shows the fit with Eq.~\ref{tempFit}. Note the data
    points above this line, indicating gas heating in small halos by
    larger scale dynamics in the vicinity of more massive halos.}
  \label{tempHalo}
\end{figure}

In the top panel of Figure~\ref{tempHalo} we plotted the mass-weighted
gas temperature

\begin{equation}
  \langle T\rangle_{\rm bin}=\int T\rho dV \bigg/ \int\rho dV
\end{equation}

\noindent
averaged over the virial spheres of all halos in a given mass bin, as
a function of that mass. Only halos containing at least 100 DM
particles are included. It is evident that a significant fraction of
the gas fails to heat to the virial temperature, since its cooling is
fairly efficient and the infall velocities are relatively small. This
result was first pointed out by \citet{binney77} and more recently by
\citet{birnboim.03}, who showed that in one-dimensional models a
virial shock does not develop in halos below some redshift-dependent
critical mass.

We find that in nested grid models lower-mass halos have the mean
temperature well above their virial temperature (Fig.~\ref{tempHalo})
indicating that they are found close to the virial radii of more
massive halos and are being heated by the gas falling into the
potential wells of their massive neighbors. A fit to all halos with
$T\lt T_{\rm vir}$, i.e. halos not affected by this ram-pressure
heating, produces a relationship

\begin{equation}
\label{tempFit}
\log\langle T\rangle_{\rm bin}=-3.51+0.80\log\mhalo,
\end{equation}

\noindent
which has a steeper slope than that of the virial temperature
($2/3$). This result can be also demonstrated by the mass fraction
$f_{\rm c}$ of gas at $T\lt10^4{\rm\,K}$ (Fig.~\ref{tempHalo}). Due to
shock heating this fraction decreases with increasing halo mass, but
in lower-mass ($10^9\lsim\mhalo\lsim10^{11}\msun$) systems most gas
can be actually found in the cold phase. As expected, the fraction
$f_{\rm c}$ traces the amount of neutral gas in halos, as can be seen
from the lower panel of Figure~\ref{tempHalo}, where we plotted the
ratio of the mass-weighted neutral fraction

\begin{equation}
  \langle x_{\rm neu}\rangle_{\rm bin}=\int{n_{\rm HI}\rho dV\over n_{\rm
      HI}+n_{\rm HII}} \bigg /\int\rho dV.
\end{equation}

\noindent
to $f_{\rm c}$. In order to explain DLA kinematics with large velocity
widths of low-ionization species, we need a sufficiently large cold
fraction $f_{\rm c}$ of gas in massive halos which have a high
velocity dispersion. On the other hand, it is expected that halos
above a certain mass scale may not contribute significantly to the
observed neutral gas absorption due to shock heating of their baryons
\citep{tinker.07}. Therefore, our goal is to determine precisely the
fraction of cold gas inside compact clouds and/or accretion filaments
which survive shock heating while falling into the massive halos. To
have a large velocity dispersion, we need to maximize the number of
such clouds intersected by any single line of sight going through the
halo. In our simulations as much as $\sim20\%$ of all gas in
$10^{12}-10^{12.5}\msun$ halos is at $T\lt10^4{\rm\,K}$
(Fig.~\ref{tempHalo}), however, this amount of cold material does not
automatically produce the observed velocity dispersion.


\begin{figure*}
  \epsscale{1.}
  \plotone{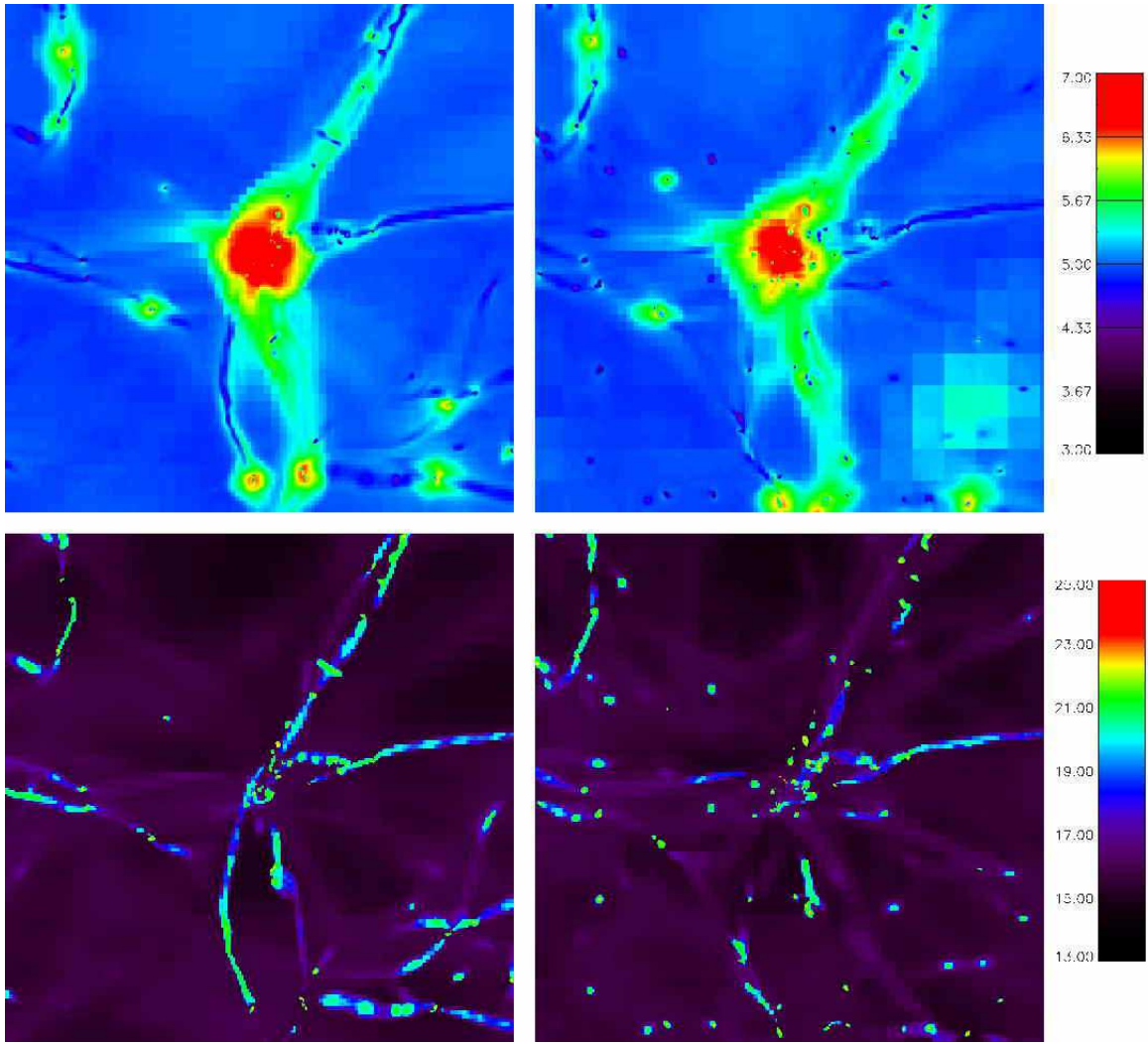}
  \caption{Mass-weighted temperature projection (top panels, in K) and
    HI column density (lower panels, in $\cmsq$) for the most massive
    halo ($8\times10^{12}\msun$) in models H1 (refined to the same
    $l_{\rm max}=7$ everywhere in the volume; left panels) and HN (the
    halo at the center was resimulated at 4X the highest grid and 64X
    the highest mass resolution of H1; right panels). Each panel is
    $4h^{-1}\mpc$ (comoving) on a side and $4h^{-1}\mpc$ thick.}
  \label{h1-nh-zoom}
\end{figure*}

\begin{figure}
  \epsscale{1.}
  \plotone{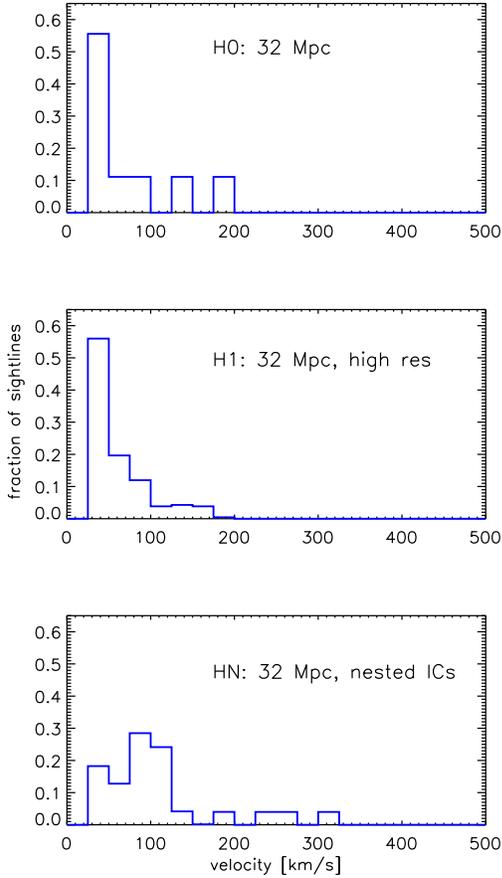}
  \caption{Distribution of velocity widths along random sightlines
    crossing the most massive halo ($8\times10^{12}\msun$) in
    $32h^{-1}\mpc$ models, at three different resolutions.}
  \label{oneHaloVelocity}
\end{figure}

One can argue that the spatial distribution of neutral clouds inside
collapsed halos is equally important for probing the velocity
space. Could improved numerical resolution change this distribution?
Without running higher resolution models of isolated galaxies and
understanding the physics of feedback on the scale of typical neutral
structures in the ISM of absorber galaxies, it is very difficult to
extrapolate our results to higher resolution. Even in our current
models resolution effects show up in several different ways. First of
all, in more massive ($\gsim10^{11}\msun$) halos high mass resolution
assists early collapse onto a larger number of low-mass DM clumps, as
opposed to slower accretion via filamentary flows onto more massive
and more centrally concentrated DM halos at somewhat lower redshifts
(Fig.~\ref{h1-nh-zoom}). Higher grid resolution leads to more
efficient cooling and in general more compact shock heated regions,
resulting in a higher survival probability of neutral clouds falling
on massive virialized halos. The combined effect of higher mass and
grid resolution in massive environments is the increased line-of-sight
velocity dispersion in a larger number of compact ($\lsim10\kpc$)
neutral gas clouds, as several of these clouds can be normally
intersected by a line of sight going through such a halo. The
resulting spectra feature multiple components and the combined
velocity widths of up to $300\kms$ (Fig.~\ref{oneHaloVelocity}), with
a median $\v90$ value close to $90\kms$ (Fig.~\ref{velMass}).

On the other hand, in lower-mass environments, such as isolated
$\lsim10^{10}\msun$ halos and filaments, higher numerical resolution
yields a more numerous halo population, and consequently, many more
neutral clouds. These clouds are sufficiently isolated so that very
few of them fall on the same line of sight, and the velocity
dispersion in individual clouds is too small to contribute to the
large observed velocity tail, independently of resolution. However,
these lower-mass environments supply the majority of low $\v90$
absorbers and are an important contributor to the overall column
density distribution.




Ideally, to combine the benefits of models of different volumes and
resolutions, one would like to convolve the incidence rates in
Figure~\ref{incidenceMass} with the velocity distributions from
Figure~\ref{velMass}, using the best resolved halos in each mass bin
in both distributions and adding intergalactic DLAs. Note that our
$4-8\mpc$ models represent fairly small volumes and are therefore
subject to the large cosmic variance. Unfortunately, these small
volumes are also the most expensive to compute, as they start to
resolve the ISM, and the fraction of the volume which is refined
adaptively rises steeply.

\section{SPH models with efficient feedback}
\label{sphModels}

We now turn to high-resolution galaxy formation models computed with
the TreeSPH code \citep{sommer-larsen..03} which, unlike Enzo, uses
particle representation of the fluid. SF is similarly modeled with a
set of discrete star particles which represent a population of stars
born at the same time. Two distinct modes of SF depending on the
thermal history of the star-forming gas are used: (1) early and very
efficient ($\epsilon_{\rm sf}=1$) SF from gas which has always been
cooler than $T_{\rm cr}=3\times10^4{\rm\,K}$ and is found above a
critical density $n_{\rm H,cr}=0.3{\rm\,cm^{-3}}$, and (2) a late mode
with much lower efficiency ($\epsilon_{\rm sf}=5\times10^{-3}$) fueled
by gas which has been heated to temperatures above $T_{\rm cr}$ and
has subsequently cooled down to be found in clouds with density above
$n_{\rm H,cr}=0.01{\rm\,cm^{-3}}$.

For our purposes the most important difference between the two codes
lies in implementation of feedback. In Enzo feedback energy is
injected into the thermal energy of the ISM which is then allowed to
be radiated away on a cooling timescale resulting in a very
conservative realization of feedback. In the TreeSPH code feedback is
modeled in accordance with a given initial mass function (IMF),
however, cooling is suppressed in the feedback regions for the
duration of the starburst, resulting in a much more efficient
expansion of the heated gas into the surrounding medium. This latter
approach has been shown to preserve a larger fraction of angular
momentum in the baryonic component as it settles into rotationally
supported disks, and to lead to more realistic disks at $z=0$
\citep{sommer-larsen..03}. Can these state-of-the-art models reproduce
the observed DLA kinematics at $z=3$?

To address this question, we selected three halos from a dark
matter-only run of box length $10h^{-1}\mpc$ which was then
resimulated with the TreeSPH code at higher resolution in Lagrangian
regions enclosing the galaxies, with gas (and star) particle masses of
$1.1\times10^6\msun$. Two of these halos (15 and 18) would form
massive galaxies at $z=0$, and the third one (Scl54) would form a
cluster. At $z=2.95$ their virial masses are $4.2\times10^{11}\msun$,
$3.0\times10^{11}\msun$ and $1.6\times10^{13}\msun$, respectively. In
the latter proto-cluster run SF and feedback is modeled with a set of
discrete star particles assuming a Salpeter IMF. One of the
proto-galaxies (AY18) was modeled with a top-heavy Arimoto-Yoshii IMF,
and the other proto-galaxy was rerun twice, once with the Salpeter
(S15) and once with the Arimoto-Yoshii (AY15) IMFs, respectively. We
then took these halos at $z=2.95$ and projected them onto a hierarchy
of nested grids with interpolated grid resolution of $45\pc$ for the
two proto-galaxies and $150\pc$ for the proto-cluster. We
postprocessed these datasets with the UVB radiative transfer using the
same algorithm as for our grid-based simulations (Sec.~\ref{rt}), and
with transfer of stellar ionizing photons using the adaptive ray
splitting method described in \citet{razoumov.06}. The stellar UV
luminosity is determined using the population synthesis package
Starburst 1999 \citep{leitherer........99} with continuous SF
distributed among all stars younger than $34\Myrs$.

\begin{figure}
  \epsscale{1.1}
  \plotone{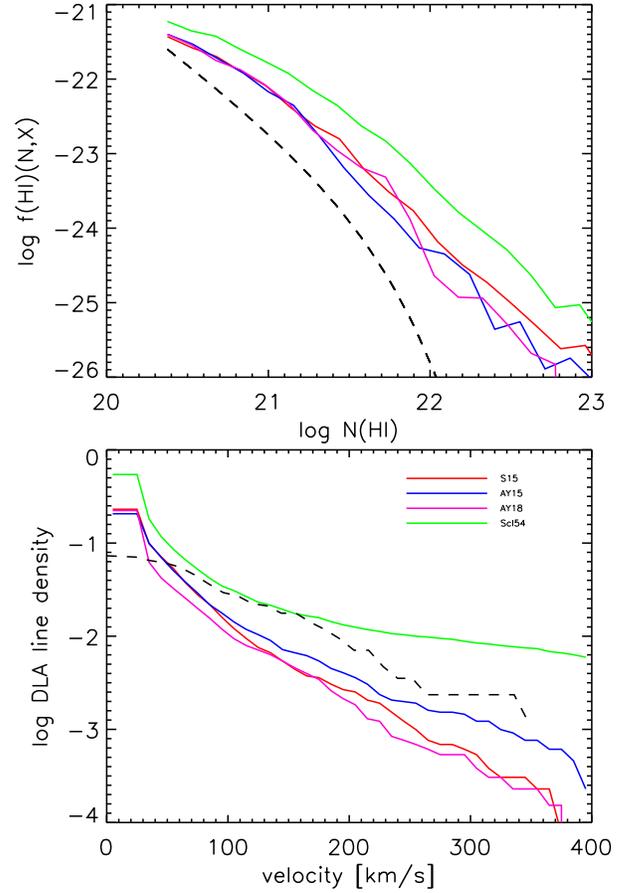}
  \caption{HI column density frequency distribution $f(N,X)$ (top
    panels) and line density $\ell_{\rm DLA}(X)$ of DLAs with the Si
    II velocity width higher than $v_{\rm Si II}$ vs. $v_{\rm Si II}$
    (lower panels) for the two Salpeter (S15 and Scl54) and two
    Arimoto-Yoshii (AY15 and AY18) models computed with the TreeSPH
    code.}
  \label{colVel-jesper-2.95}
\end{figure}

Figure~\ref{colVel-jesper-2.95} shows the differential HI column
density and cumulative $\v90$ velocity width distributions for the
four models. These distributions were computed with the random
sightlines passing through the resimulated volumes which have the
average densities in excess of the mean density of the Universe at
$z=2.95$. Therefore we should expect more absorption than from a truly
representative large cosmological volume, which is evident in the
upper plot of Fig.~\ref{colVel-jesper-2.95}. However, since these
volumes contain massive environments, they should exhibit a higher
neutral gas velocity dispersion than a larger cosmological volume. The
median $\v90$ velocity width for models S15, AY15, AY18, and Scl54 is
32, 34, 31, and $31\kms$, respectively. This is a surprising result,
as with suppression of cooling in the feedback regions one would
expect to see a more efficient conversion of feedback energy into
hydrodynamical expansion and, therefore, higher outflow velocities. In
fact, we see that increasing the strength of feedback has some effect
on velocities. AY15 is the same proto-galaxy as S15, resimulated with
the top-heavy Arimoto-Yoshii IMF with stronger
feedback. Figure~\ref{colVel-jesper-2.95} shows that more energetic
feedback removes gas from higher column densities and produces a
larger fraction of high-velocity DLAs, although the median $\v90$
changes only slightly from 32 to $34\kms$.

Note that our kinematics results are independent of the assumed
UVB. Changing the amplitude and hardness of the UVB does not alleviate
the lack of high-velocity systems, as it affects primarily lower
column density absorbers, and since there is effectively no
correlation between the velocity widths and column densities (lower
panel of Fig.~\ref{scatter-colVel-boxSize}), changing the UVB just
adds or removes both low- and high-velocity DLAs from the simulated
sample.

\section{Discussion}

We have performed a numerical study of DLA kinematics using
high-resolution grid-based AMR simulations with moderate feedback, as
well as particle-based SPH models with much more efficient
feedback. With grid-based models our goal was to test the hypothesis
that the observed velocity dispersion in DLAs could be explained by
energy coming from gravitational collapse of cosmic structures. The
broadest line profiles would then come from neutral clouds surviving
infall into the massive virialized ($\gsim10^{11}\msun$) systems, as a
$10^{11}\msun$ halo at $z=3$ has a circular velocity already in excess
of $100\kms$. We used a series of models with a comoving volume size
ranging from $4h^{-1}$ to $32h^{-1}\mpc$ and the maximum physical
resolution of $100\pc$ to build a DLA population covering halos in the
mass range $10^{8.5}-10^{13}\msun$. Although our models can reproduce
well the total incidence rate of DLAs at $z=3$, our median $\v90$
velocities are of order $40-50\kms$, well below the observed value of
$90\kms$. We can point out three possibilities to resolve this
discrepancy:

1. {\it More massive environments --} The velocity dispersion in
individual clouds is a strong function of mass. Therefore, one could
expect that including more massive halos through sampling of the rare
density peaks could alleviate the velocity problem. On the other hand,
all our 4-16 Mpc high resolution models feature similar velocity tails
(Fig.~\ref{colVel-boxSize}) even though the mass function in the
$16\mpc$ model L1 extends to nearly 50 times more massive halos than
in the $4\mpc$ model N1. Therefore it seems unlikely that the more
massive ($\mhalo\gt8\times10^{12}\msun$) and consequently much rarer
halos could make any substantial contribution to the overall
statistics, even though each individual halo could have a velocity
dispersion of several hundred $\kms$.

Note that including only massive halos in larger simulation volumes
might produce $f(N,X)$ at low $\nhi$ which is too small
\citep{jedamzik.98} or too large, depending on the technique. In our
simulations neglecting low-mass halos can actually raise $f(N,X)$ at
low column densities, as all the gas which would otherwise be pulled
in by low-mass halos stays in extended filaments which sometimes can
produce column densities $\nhi\gt10^{20.3}\cmsq$. On the other hand,
our larger $32\mpc$ models H0 and H1 do not have enough grid
resolution to resolve these filaments, and hence produce small line
densities $\ell_{\rm DLA}(X)$.

2. {\it Resolution --} Our current simulations are already very
expensive computationally as we refine adaptively everywhere in the
volume, and the $128^3$ base grid models have in practice $300^3$ to
$400^3$ resolution elements. If we start from $256^3$ or a larger base
grid, many more halos will form drawing in a larger fraction of
baryons at earlier redshifts, whereas inside individual galaxies we
will start resolving clumpy interstellar gas. However, without running
these simulations and sampling a large number of galaxies at higher
resolution, it is impossible to predict reliably the effect on the
velocity and column density distributions. In addition, higher
resolution would allow us to better model propagation of
supernova-driven winds in the clumpy interstellar gas.

The key question is the relative contribution of lower mass
($\lsim10^{10.5}\msun$) and more massive ($\gsim10^{10.5}\msun$)
halos. In our current models these two populations contribute about
equally to the total ``galactic'' DLA column density
(Fig.~\ref{incidenceMass}). It is possible that at higher resolution
the relative contribution of the more massive halos might grow, as
they will feature more numerous spatially separated components falling
on the same line of sight, perhaps with neutral gas tidal tails and
bridges, whereas isolated lower mass galaxies could become even more
compact and pull the remaining cold neutral gas from extended
filamentary structures decreasing its covering factor on the sky.

Unfortunately, the spatial distribution of absorbing clouds cannot be
reconstructed from the observed line profiles in the velocity space,
which would otherwise point to the minimum scale needed to resolve
muli-component DLAs. \citet{lopez...02} studied an almost dust free
DLA at $z=2.33$ featuring 14 resolved metal line components. They
found very similar Z/Fe ratios of low-ionization species in all 14
components, suggesting that these components could originate in clouds
physically located in a small region of space, perhaps much smaller
than the resolution limit of our current simulations.

3. {\it Local microphysics --} The most intriguing possibility is that
the observed velocity dispersion could arise as a result of feedback
from SF. More efficient feedback could disrupt the highest column
density absorbers ($\nhi\gsim10^{21.5}\cmsq$) in the lower panel of
Fig.~\ref{scatter-colVel-boxSize} creating a population of low
$\nhi$ systems with high velocity widths. With such a disruption
mechanism put in by hand with the galactic wind model,
\citet{nagamine...07} saw efficient gas removal from lower-mass
($\lsim10^{10}\msun$) halos and increased cross-sections in
$\gsim10^{11}\msun$ halos. In view of our results, a model in which
more efficient feedback is obtained without resorting to the
unphysical assumptions of kinematic winds and/or suppressing cooling
in feedback regions would seem particularly compelling. The key
challenge here is to come up with a model which describes dynamics of
individual components of the multiphase ISM separately from each
other, either resolving them explicitly, or using a subresolution
model in which different components are advected differently on a
grid. In these models supernova winds would channel most of their
energy into the lower density regions between the star-forming clouds,
whereas the clouds would be more likely to survive the disruptive
effect of the winds.

In Section~\ref{sphModels} we used TreeSPH models to test the effect
that the suppression of cooling has on neutral gas kinematics. We
found the median velocity widths $\v90$ of only 31 to $34\kms$, even
though that these models have been shown to expel gas efficiently from
the star-forming regions at high redshifts and produce realistic
galactic disks at $z=0$. In our view, such low velocity dispersion
could be explained by a number of factors.

A. {\it Resolution --} When numerical models do not resolve density
inhomogeneities in the ISM, putting a certain amount of supernova
feedback into the thermal energy of the gas and turning off cooling
leads to an efficient conversion of this energy into the superwind
expansion into a nearly uniform medium. However, the mass loading of
the wind in a uniform medium is much larger than in a clumpy medium of
the same average density, since in the latter case the wind will
predominantly expand into the lower density voids between the clumps
creating a high velocity outflow which will occasionally sweep chunks
of neutral material off the edges of denser clouds. Clumps of gas
entrained in winds are often observed in low-redshift outflows
\citet{rupke..05}. On the other hand, at low resolution winds will
displace most gas in the galaxy moving it to somewhat higher galactic
radii, producing ``puffy'' galaxies with a relatively low velocity
dispersion. This is exactly what we see in SPH simulations at $45\pc$
physical resolution. The expectation is that at higher resolution we
should see multiphase outflows, with higher overall velocities in hot
ionized gas, and numerous dense clouds giving rise to multi-component
low-ionization absorption lines. Getting such galactic winds from
first principles is notoriously difficult in numerical simulations. In
this paper, we argue that the necessary condition for obtaining the
correct DLA velocity dispersion is a separate dynamical treatment of
different components of the multiphase interstellar medium at $z=3$.

In addition to the resolution effects, two other factors could have
affected our SPH results. Here we just mention these effects briefly,
as clearly their study goes beyond the reach of this paper.

B. {\it Mean redshift of feedback --} A realistic population of
galaxies at $z=0$ can be obtained in cosmological simulations invoking
early ($z\gsim 4-6$) episodes of star formation with energetic
feedback \citep{sommer-larsen..03}. In this scenario, the radial
distribution of gas and its angular momentum arise as a cumulative
result of successive episodes of star formation driving gas out of the
galaxies. Other model parameters might result in similar star
formation histories; not all parameters can be constrained uniquely
from observations. One could speculate that a different star formation
history, e.g. one featuring energetic feedback down to redshift
$z\sim3$, would produce a higher neutral hydrogen velocity dispersion
at such a redshift. We note that from the spectra of $z\sim3-4$ LBGs,
outflow velocities of $300-400 \kms$ are routinely inferred
\citep[e.g.,][]{pettini.......01, shapley...03}. Note, however, also
that \citet{laursen.07} were able to match both the magnitude and
radial fall-off of Ly$\alpha$ surface brightness of typical $z\sim3$
galaxies, indicating that the spatial distribution of neutral hydrogen
is correctly predicted by current SPH models.

C. {\it AGN feedback --} It is possible that the core of each massive
galaxy hosts a supermassive black hole which might have shown an
AGN-type activity at some time in the past. Neither grid-based nor
particle-based simulations presented in this paper include feedback
from AGNs which could have a profound impact on gas kinematics in host
galaxies.

Feedback from star formation is generally thought to play an important
role in galaxy evolution. If galactic winds give rise to the observed
neutral gas kinematics in DLAs, the same winds should affect the
column density distribution. In Paper I and in this paper, we have
attempted to look at both distributions through large-scale,
``blind-survey'' models, in which we simulate cosmological volumes
using aggressive grid refinement for all galaxies. However, these
models are very expensive to compute, especially as we get closer to
resolving the typical scales of the ISM. Perhaps, a more efficient
approach is to combine these large-scale numerical surveys with
simulations of the ISM in isolated galaxies, whether or not accounting
for mergers and cosmic infall. Fortunately, there is sufficient amount
of observational data which can be used to constrain these
models. Combining traditional QSO-DLA data with a closer look at the
high-redshift star-forming regions with GRB-DLAs and with emission
line studies should allow us to learn a lot more about young
proto-galaxies in coming years.

\acknowledgments

The authors are grateful to the anonymous referee for his/her careful
reviewing and constructive comments. The AMR simulations were
performed at the San Diego Supercomputer Center (SDSC). The TreeSPH
simulations were performed on the SGI Itanium II facility provided by
the Danish Center for Scientific Computing (DCSC). This research was
supported by the DFG cluster of excellence ``Origin and Structure of
the Universe''. The Dark Cosmology Centre is funded by the DNRF.

\bibliographystyle{apj}


\end{document}